\documentclass[aps,prb,twocolumn,groupedaddress,showpacs]{revtex4-1}

\usepackage{graphics}

\begin{document}

\title{
Improvement of magnetic hardness
at finite temperatures:\\
{\it ab initio} disordered local moment approach for YCo$_5$}

\author{Munehisa Matsumoto$^{1}$, Rudra Banerjee$^{2}$, and Julie B. Staunton$^{2}$}
\affiliation{$^1$Elements Strategy Initiative Center for Magnetic Materials (ESICMM),\\
National Institute for Materials Science, Tsukuba 305-0047, JAPAN\\
$^{2}$Department of Physics, University of Warwick, Coventry CV4 7AL, United Kingdom}

\date{\today}

\begin{abstract}
Temperature dependence
of the magnetocrystalline anisotropy energy
and
magnetization of the prototypical rare-earth magnet YCo$_5$ is calculated
from first principles, utilizing the relativistic disordered local moment approach.
We discuss a strategy to enhance the finite-temperature anisotropy field
by hole doping, paving the way for an improvement of the coercivity
near room temperature or higher. 
\end{abstract}



\maketitle

\section{Introduction}

Ferromagnetism in permanent magnets
has been a long-standing problem in solid state physics and
materials science.
The mechanism of coercivity,
resistance of the spontaneous magnetization
against the externally applied magnetic field in the reverse-direction,
is not yet entirely known after almost 70 years
since Brown's paradox~\cite{brown_1945}.
Development of modern rare-earth magnets
like Nd$_2$Fe$_{14}$B~\cite{sagawa_1984}
and SmCo$_5$~\cite{tawara_1968,nesbitt_1968}
triggered material-specific investigations of
the coercivity mechanism~\cite{kronmuller_1988}.
They culminate in today's state-of-the-art
technological optimization of the
microstructure of ferromagnetic materials for the
enhanced coercivity~\cite{hono_2012}
on the phenomenological and empirical basis.
More fundamental understanding and controlling on the basis of
solid-state physics and statistical physics is desired
to better use these technological advances.
Especially the temperature decay of coercivity
is a problem in industrial applications and thus
more and more demands for the control of
finite-temperature properties of permanent
magnets are rising.

The high-temperature tail of
the magnetism is mostly carried by $3d$-electrons~\cite{skomski_1998}
in the intermetallic system
between transition-metal (TM) and rare-earth (RE) elements
while the low-temperature magnetism is dominated by
$4f$-electrons.
Thus developing
an understanding and the
theoretical control over the temperature decay of $3d$-electron magnetism
has the universal utility for the optimization
of high-temperature magnetism of RE magnets.
Thus we focus on YCo$_5$ as our target system,
extracting the $3d$-electron part out of the SmCo$_5$ family.
The so-called 1-5 magnets represent
the RE magnets with one of the simplest crystal structures
and SmCo$_5$-based magnets
are 2$^{nd}$ in magnetic performance
only
to the champion-magnet family
including Nd$_{2}$Fe$_{14}$B; SmCo$_5$ currently
serves for some special applications
under extreme conditions~\cite{ohashi_2012}
with its high Curie temperature and stability against corrosion.
Also the large magneto-crystalline anisotropy energy (MAE)
of 5.5~[MJ/m$^3$]
at room temperature $T=295~\mbox{[K]}$ in YCo$_5$~\cite{strnat_1967}
can let itself qualify for permanent magnet applications~\cite{ohashi_2012}.

Describing the magnetic properties of such a material presents significant 
challenges to {\it ab initio} theory.
Itinerant electron magnetism
in realistic magnetic materials
has often
been addressed in the
ground state with the application of the
density-functional theory (DFT). Incorporation of finite temperature physics
and description
of correlated electron
behavior can be issues for DFT
which does not always include
localized electron effects adequately into its scope. Both issues require a combination of DFT and statistical physics in some way.
Here we follow the idea of disordered local moments~\cite{hubbard_1979,hasegawa_1979,hubbard_1979_II,hubbard_1981}
to incorporate the physics of thermal fluctuations of well-developed
magnetic moments
into the spin-polarized density functional theory (DFT) for
ferromagnetic materials~\cite{pindor_1983,gyorffy_1985,julie_1985,julie_1986}.
A relativistic version of the disordered local moment (DLM) approach has recently been
developed~\cite{julie_2004, julie_2006} and successfully applied to finite-temperature
magnetic anisotropy in
L1$_{0}$-FePt alloys~\cite{julie_2004, julie_2006} and
Co films~\cite{buruzs_2007} where magnetism is all carried by $d$-electrons.
Thus the $d$-electron part in the magnetism
of rare-earth magnets can be a good target for DLM.
\begin{figure}
\begin{center}
\scalebox{0.8}{\includegraphics{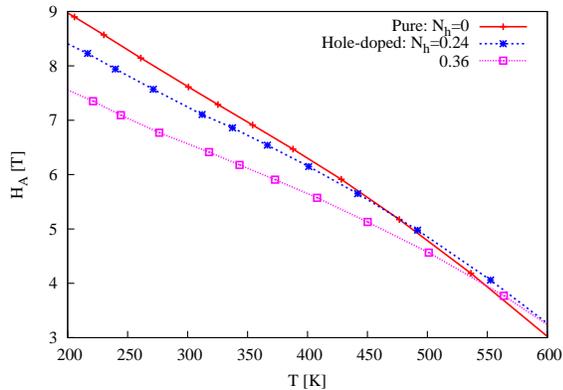}}\\
\end{center}
\caption{\label{main_message} (Color online) Calculated temperature dependence of
the anisotropy field $H_{\rm A}$ for pure and hole-doped YCo$_5$,
where
doping-induced enhancement of $H_{\rm A}$ is observed
in the temperature range $T\stackrel{>}{\sim} 450$~[K]
and $550$~[K]
with the number of
doped holes being $N_{h}=0.24$ and $0.36$, respectively.
}
\end{figure}

In this work we present {\it ab initio} calculations based on DLM
for YCo$_5$ at finite temperatures.
From calculated temperature dependence of the magnetization $M$ and
the MAE, which we denote by $K_{\rm u1}$ as it mostly comes from the uni-axial
magneto-crystalline anisotropy (MCA) for YCo$_5$, we
address the thermal decay of the anisotropy field
estimated as $H_{\rm A}=2K_{\rm u1}/M$
which
generally shows the approximate
proportionality to
the coercivity field $H_{\rm c}$~\cite{hirosawa_1986_jmmm}.
The relation is practically $H_{\rm c} \sim 0.2 H_{\rm A}$
with the practical hard upper limit
being $H_{\rm c}\stackrel{<}{\sim}0.5 H_{\rm A}$ that is
reached only in some exceptional samples~\cite{hono_hirosawa_2012}.
Assuming such rough proportionality, we propose
a strategy to improve $H_{\rm A}(T)$ at high temperatures in the
range $T\sim 400-500~\mbox{[K]}$. Our main message
is encapsulated in Fig.~\ref{main_message} which
shows that
hole-doped YCo$_5$ with the number of doped holes being 0.24
and 0.35
per unit cell
has the stronger anisotropy field than pure YCo$_5$ in the temperature range
$T\stackrel{>}{\sim}450~\mbox{[K]}$ and $T\stackrel{>}{\sim}550~\mbox{[K]}$, respectively.

\begin{figure}
\begin{center}
\scalebox{0.8}{\includegraphics{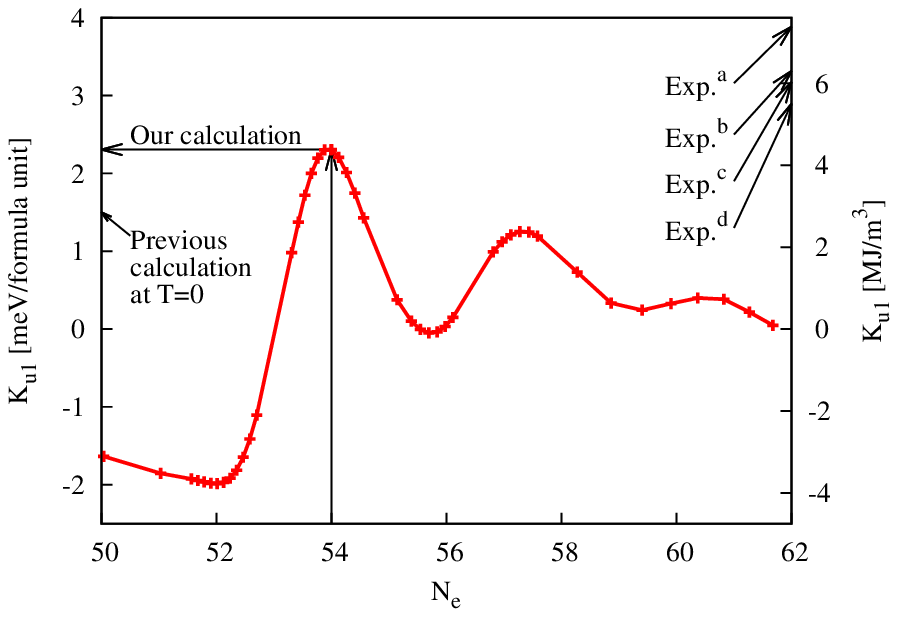}}
\end{center}
$^a$ $K_{\rm u1}=7.38~\mbox{[MJ/m$^3$]}$ at $T=4.2$~[K]
from Ref.~\onlinecite{givord_1981}.\\
$^b$ $K_{\rm u1}=6.3~\mbox{[MJ/m$^3$]}$ at $T=77$~[K] from Ref.~\onlinecite{buschow_1974}.\\
$^c$ $K=6.03~\mbox{[MJ/m$^3$]}$ from Ref.~\onlinecite{Pr_doping_1988} at $T=293$~[K]. We assume $K_{\rm u1}$ dominates in $K$ as was shown in Ref.~\onlinecite{givord_1981}.\\
$^d$ $K_{\rm u1}=5.5~\mbox{[MJ/m$^3$]}$ at room temperature
from Ref.~\onlinecite{ohashi_2012}.
\caption{\label{T_0_pure} (Color online)
Calculated magnetic anisotropy energy
plotted as the function of the valence electron number
for YCo$_5$
near the ground state $T=0$~[K].
The numbers on the left-hand side of the vertical axis
indicates the results per the formula unit of YCo$_5$
and those on
the right-hand side of the vertical axis gives
the same quantities
converted to per the unit volume
using the experimental
lattice constants in Ref.~\onlinecite{neutron_1980}.
The  previous theoretical result for MAE at $T=0$
is taken from Ref.~\onlinecite{igor_2003}.
The number of valence electrons per formula unit of YCo$_5$ is 54.
}
\end{figure}
Experimentally, doping-enhanced coercivity has been known
for YCo$_{5-x}$Ni$_{x}$ and other RECo$_{5-x}$Cu$_{x}$
magnets (RE=Ce, Sm)~\cite{buschow_1976}.
The coercivity is indeed
significantly affected by doping,
which is seen in our calculated results on
the filling dependence of MAE near the ground state as shown in Fig.~\ref{T_0_pure}.
On top of such
a filling dependence of coercivity determined by the electronic structure,
our idea is to let the electronic states
below
the peak of MAE
in Fig.~\ref{T_0_pure}
be thermally populated so that the high-temperature MAE
is enhanced.
Experimentally observed doping-enhanced coercivity
had been discussed in conjunction with
intrinsic pinning~\cite{buschow_1976}.
Our results
on the finite-temperature
physics together with the
electron band filling-sensitive nature of MAE
point to a new scenario for
doping-enhanced coercivity at high temperatures.

The rest of paper is organized as follows.
Our proposal on the strategy to enhance the high-temperature
coercivity is found in Sec.~\ref{sec::coercivity}.
This follows an outline of the DLM 
methodology underpinning the calculations in Sec.~\ref{sec::methods}
and Sec.~\ref{pure}.
Conclusions are given in Sec.~\ref{conc}.

\section{Methods}
\label{sec::methods}

Our computational method, the DLM approach, is
based on the Korringa-Kohn-Rostoker (KKR) method~\cite{KKR,KKR2}
and
coherent potential approximation (CPA)~\cite{soven_1967}
for {\it ab initio} electronic structure calculations.
Especially to address the temperature dependence of MAE
which is brought about by spin-orbit interaction (SOI),
we follow the relativistic formulation~\cite{julie_2004, julie_2006}.
These aspects in
Sec.~\ref{method::framework} and Sec.~\ref{method::MAE},
describe the general framework and how to extract MAE, respectively
at the level of sketching the basic ideas. The MAE as our
key observable
is
focused on and
for the rest of the observables
that we present in Sec.~\ref{pure},
we refer to the original works~\cite{gyorffy_1985,julie_2006}
and a recent review article~\cite{ebert_2011} for their explicit expressions
in terms of KKR-CPA. In Sec~\ref{method::details}
we give the details of the application
to the case of YCo$_5$.

\subsection{Framework}
\label{method::framework}

\subsubsection{Separation of fast and slow dynamics}
\label{separating_dynamics}

Under the situation that
spin-fluctuation dynamics are much slower than the characteristic time scales
of electronic motions~\cite{gyorffy_1985},
well developed classical local moments $\hat{\mathbf{e}}$
are assumed to exist on each
site $i=1\ldots N$ in the unit cell,
with $N$ being the number of magnetic
atoms in the unit cell.
The thermal average $\left<\right>$ of the local moment
\begin{equation}
\left<
\hat{\mathbf e}_{i}
\right>=\int\cdots\int\hat{\mathbf e}_{i}P^{(\hat{n})}(\{\hat{\mathbf e}\})
d\,\hat{\mathbf e}_{1}\ldots\,d\,\hat{\mathbf e}_{N}
\label{average_moment}
\end{equation}
is assumed to be
aligned with the magnetization direction $\hat{\mathbf n}$ for a ferromagnet. Here $\hat{\mathbf n}$ is
fixed in the calculation by hand. We have
specified the configuration of the local moments
by $\{\hat{\mathbf e}\}$
and
statistical weight 
$P^{(\hat{n})}(\{\hat{\mathbf e}\})$
given in terms of the Boltzmann weights
at finite temperatures.
$
\left<
\hat{\mathbf e}_{i}
\right>=\mathbf{m}_{i}
$
is a magnetic order parameter.
For a ferromagnet, magnetized along a direction $\hat{\mathbf n}$, 
$\mathbf{m}_{i} = m_i \hat{\mathbf n}$.

\subsubsection{DLM as a realistic mean field theory}

DLM is implemented as a realistic mean-field theory
for
the local moments
embedded in a sea of electrons
given by spin-polarized DFT for the target material.
For a given configuration of the local moments
\[
\{\hat{\mathbf e}\}\equiv
\{
\hat{\mathbf e}_{1},
\hat{\mathbf e}_{2},
\ldots,
\hat{\mathbf e}_{N}
\}
,
\]
the Boltzmann weight is defined as
\[
\exp\left[
-\beta\Omega^{(\hat{\mathbf n})}(\{\hat{\mathbf e}\})
\right],
\]
referring to the grand potential $\Omega^{(\hat{\mathbf n})}(\{\hat{\mathbf e}\})$
in the spin-resolved density functional theory for a given target ferromagnet
($\beta= 1/k_B T$).
The partition function is accordingly written as
\[
Z^{(\hat{\mathbf n})}=\int\cdots\int
\exp\left[
-\beta\Omega^{(\hat{\mathbf n})}(\{\hat{\mathbf e}\})
\right]
d\,\hat{\mathbf e}_{1}\ldots\,d\,\hat{\mathbf e}_{N}
\]
and the probability distribution function (PDF)
in Eq.~(\ref{average_moment})
of the local-moment
configuration $\{\hat{\mathbf e}\}$ is
\[
P^{(\hat{\mathbf n})}=\frac{
\exp\left[
-\beta\Omega^{(\hat{\mathbf n})}(\{\hat{\mathbf e}\})
\right]
}
{
Z^{(\hat{\mathbf n})}
}
\]

Here we use the mean-field approximation (MFA)
\begin{equation}
\Omega_{0}^{(\hat{\mathbf n})}(\{\hat{\mathbf e}\})
\stackrel{\rm MFA}{=}
\sum_{i=1}^{N}
{\mathbf h}_{i}^{(\hat{\mathbf n})}\cdot \hat{\mathbf e}_{i}
\label{MFA}
\end{equation}
where
the
Weiss field ${\mathbf h}_{i}^{(\hat{\mathbf n})}=h_{i}^{(\hat{\mathbf n})}\hat{\mathbf n}$
is written as follows~\cite{gyorffy_1985}.
\begin{equation}
h_{i}^{(\hat{\mathbf n})}=\int\frac{3}{4\pi}
\left(
\hat{\mathbf e}_{i}
\cdot
\hat{\mathbf n}
\right)
\left<
\Omega^{(\hat{\mathbf n})}
\right>_{\hat{\mathbf e}_{i}}
d\,\hat{\mathbf e}_{i}
\label{Weiss_field_expressed_as_}
\end{equation}
for $i=1\ldots N$ where $i$ runs over the local moments in the unit cell
and the thermal average denoted by
$\left< \cdot \right>_{\hat{\mathbf e}_{i}}$
implies the restriction in the averaging process
with the direction of the local moment
on site $i$ fixed to that specified by $\hat{\mathbf e}_{i}$.
(Further terms proportional to, say, $(\hat{\mathbf e}_i \cdot \hat{\mathbf n})^2$, can 
be added to Eq.~\ref{MFA} to improve the mean field description~\cite{1403.2904} but have a rather small
effect.)
With the MFA in Eq.~(\ref{MFA}) the overall
partition function factorizes into contributions from
each of local moments
\[Z^{(\hat{\mathbf n})}=\prod_{i=1}^{N}Z^{(\hat{\mathbf n})}_{i}\] and
the partition function is written in terms of the Weiss field
\begin{eqnarray}
Z^{(\hat{\mathbf n})}_{i}&=&
\int
\exp\left(
-\beta
{\mathbf h}^{(\hat{\mathbf n})}_{i}\cdot
\hat{\mathbf e}_{i}
\right)
d\,\hat{\mathbf e}_{i} \nonumber\\
&=&
\frac{4\pi}{\beta h^{(\hat{\mathbf n})}_{i}}
\sinh \beta
h^{(\hat{\mathbf n})}_{i}
\end{eqnarray}
and 
the PDF
is written as follows:
\begin{eqnarray}
P^{(\hat{\mathbf n})}_{i}(\hat{\mathbf e}_{i})
&=&\exp\left(
-\beta{\mathbf h}_{i}^{(\hat{\mathbf n})}\cdot\hat{\mathbf e}_{i}\right)
/Z^{(\hat{\mathbf n})}_{i} \nonumber
\\
&=&
\frac{\beta h_{i}^{(\hat{\mathbf n})}}
{4\pi\sinh\beta h_{i}^{(\hat{\mathbf n})}}\exp
\left(
-\beta
{\mathbf h}^{(\hat{\mathbf n})}_{i}\cdot
\hat{\mathbf e}_{i}
\right)
\label{PDF}
\end{eqnarray}
With this expression for the PDF,
the free energy
is written as follows:
\begin{eqnarray}
F^{(\hat{\mathbf n})}&=&
\left<\Omega^{(\hat{\mathbf n})}
\right> \nonumber \\
&&+\frac{1}{\beta}
\sum_{i=1}^{N}
\int
P_{i}^{(\hat{\mathbf n})}(\hat{\mathbf e}_{i})
\log
P_{i}^{(\hat{\mathbf n})}(\hat{\mathbf e}_{i})
d\,\hat{\mathbf e}_{i} \label{eq::free_energy}
\end{eqnarray}
where the first term describes the internal energy and the second describes
$-TS$ with the magnetic entropy
\[
S=-\sum_{i=1}^{N}
\int
P_{i}^{(\hat{\mathbf n})}(\hat{\mathbf e}_{i})
\log
P_{i}^{(\hat{\mathbf n})}(\hat{\mathbf e}_{i})
d\,\hat{\mathbf e}_{i}.
\]

The magnetization is given using the PDF in Eq.~(\ref{PDF})
as follows.
\begin{eqnarray}
{\mathbf m}_{i}^{(\hat{\mathbf n})}& = &
\left<
\hat{\mathbf e}_{i}
\right>
\equiv
m^{(\hat{\mathbf{n}})}_{i}
\mathbf{h}^{(\hat{\mathbf{n}})}_{i}
\\
m^{(\hat{\mathbf n})}_{i}
&=&
-\frac{1}{\beta
h^{(\hat{\mathbf n})}_{i}}
+\coth
\beta
h^{(\hat{\mathbf n})}_{i}=
L(
\beta
h^{(\hat{\mathbf n})}_{i}
)\label{Langevin}
\end{eqnarray}
Here $L(x)$ is
the Langevin function.

\subsubsection{CPA to embed the thermal disorder into DFT}

Coherent Potential Approximation (CPA)~\cite{soven_1967}
is an approach to deal with the disorder physics
on a mean-field level by averaging over the
random potentials to introduce an effective uniformly distributed
potential. Following the original idea by
Hasegawa~\cite{hasegawa_1979} and Hubbard~\cite{hubbard_1979_II},
thermal disorder in local moments is embedded
into the description of the electronic states~\cite{gyorffy_1985} 
on the level of CPA.

The way that the CPA determines the effective medium is
by letting the motion of an electron simulate the motion of an electron on the average.
The scattering problem at the heart of KKR
is solved
to get the single-site $t$-matrix 
$\underline{t}_{i}$
for each local moment $i$
where the underlined $t$-matrix is that in
orbital and spin angular momentum space.
Some more elaboration on that solution
is given below with Eq.~(\ref{rotating_t_matrix}).
For the system magnetized along the direction $\hat{\mathbf n}$,
the medium is specified
by a set of CPA-imposed
$t$-matrices~\cite{prl_1978}
\[
\underline{\underline{t}}_{c}^{(\hat{\mathbf n})}
\equiv
\left(
\underline{t}_{1,c},
\underline{t}_{2,c},
\ldots
\underline{t}_{N,c}
\right)
\]
which is determined (for details we refer to Ref.~\onlinecite{julie_2006})
for a given set of Weiss fields
$h_{i}^{(\hat{\mathbf n})}$
as in Eq.~(\ref{Weiss_field_expressed_as_}) and
corresponding
$P^{(\hat{\mathbf n})}_{i}(\hat{\mathbf e}_{i})$
as in Eq.~(\ref{PDF}), where the finite-temperature physics
is encapsulated.

The $t$-matrices $\underline{\underline{t}}_{c}^{(\hat{\mathbf n})}$ that specify
the effective medium comes into the formulation of DFT as follows.
By magnetic force theorem
we consider only the single-particle energy
part of the spin-resolved DFT grand potential
as our effective local-moment Hamiltonian
\begin{equation}
\Omega^{(\hat{\mathbf n})}(\{\hat{\mathbf e}\})
\simeq
-\int\,d\,E\,
f_{\rm FD}(E; \mu^{(\hat{\mathbf{n}})})
N^{(\hat{\mathbf n})}(E; \{\hat{\mathbf e}\}),
\label{effective}
\end{equation}
where
$f_{\rm FD}(E; \mu^{(\hat{\mathbf{n}})})
$
is the Fermi-Dirac distribution function with the chemical potential
being $\mu^{(\hat{\mathbf{n}})}$ and $N^{(\hat{\mathbf n})}(E; \{\hat{\mathbf e}\})$
is the integrated density of states (DOS)~\cite{faulkner_1980,julie_2006}
with the given
direction of magnetization
$\hat{\mathbf n}$ and the given configuration of local moments $\{\hat{\mathbf e}\}$.
Using the Lloyd formula~\cite{lloyd_1975}
for
$N^{(\hat{\mathbf n})}(E; \{\hat{\mathbf e}\})$
on the right-hand side of Eq.~(\ref{effective}),
the grand potential is expressed in terms of the $t$-matrices~\cite{julie_2006}.

\subsection{Calculation of magnetic anisotropy energy}
\label{method::MAE}

\subsubsection{Defining the MAE for the uni-axial MCA}

For uniaxial magnets,
the free energy takes the form
\[
F^{(\hat{\mathbf n})}=F_{\rm iso}+K_{\rm u 1}\sin^{2}\theta
\]
where $F_{\rm iso}$ is the isotropic part,
$K_{\rm u 1}>0$ is the MAE for uni-axial MCA
and $\theta$ is the angle between the direction of magnetization
and the easy axis.
Then
\begin{eqnarray}
T_{\theta} & = & -\frac{\partial F}{\partial\theta} \label{eq::torque_as_the_derivative}\\
& = &
-2K_{\rm u 1}\sin\theta\cos\theta\nonumber
\label{eq::torque_formula}
\end{eqnarray}
Thus in order to extract MAE for the uni-axial MCA
we fix the direction
of magnetization to be $\hat{\mathbf n}=(1,0,1)/\sqrt{2}$
that is, $\hat{\mathbf n}=(\sin\theta\cos\phi,\sin\theta\sin\phi,\cos\theta)$ with $\theta=\pi/4$
and $\phi=0$
and calculate the magnetic torque $T_{\theta=\pi/4}$.
Our target MAE is obtained as
\begin{equation}
K_{\rm u 1}=-T_{\theta=\pi/4}.
\end{equation}

\subsubsection{
Spin-Orbit Interaction (SOI) in relativistic KKR}

For a given set of self-consistent potentials,
electronic charge, and local moment magnitudes,
the directions of local moments are described
by the
unitary transformation of the
single-site $t$-matrix for local moment $i$
\begin{equation}
\underline{t}_{i}(E;\hat{\mathbf e}_{i})
=\underline{R}(
\hat{\mathbf e}_{i}
)
\underline{t}_{i}(E;\hat{\mathbf z})
\underline{R}(
\hat{\mathbf e}_{i}
)^{\dag}
\label{rotating_t_matrix}
\end{equation}
where
$\underline{t}(E;\hat{\mathbf z})$ is the $t$-matrix
with the effective field pointing along the local $z$-axis
for given energy $E$, and $\underline{R}(\hat{\mathbf e}_{i})$
is for
the $O(3)$
unitary
transformation that rotates the $z$-axis along the direction of $\hat{\mathbf e}_{i}$.
The scattering problem at the heart of KKR
is solved incorporating SOI with a magnetic field pointing along
the $z$-axis~\cite{relativistic_scattering} and the single-site $t$-matrix 
$\underline{t}_{i}(E;\hat{\mathbf z})$ is obtained.

In the absence of SOI,
an element of the single site scattering 
$t$-matrix
\begin{equation}
t_{i; l,m,\sigma; l^{\prime},m^{\prime},\sigma^{\prime}}= 
t_{i;l,\sigma}\delta_{l,l^{\prime}}\delta_{m,m^{\prime}}
\end{equation}
where $l$ is the angular momentum quantum number and $m$ the azimuthal component and
\begin{equation}
\underline{t}_{i}(E;\hat{\mathbf e}_{i})= \underline{t}_{i}^{+} \underline{1} + 
\underline{t}_{i}^{-}{\mathbf \sigma} \cdot {\mathbf e}_{i}
\end{equation}
where an element $t^{+(-)}_{l,m; l^{\prime},m^{\prime}}=\frac{1}{2} 
(t_{i;l,\frac{1}{2}} + (-) t_{i;l,-\frac{1}{2}})$.
When SOI
is included,
the $t$-matrix is no longer independent of $m$ which
leads to the generation of magnetic anisotropic effects.

\subsubsection{Torque-based formula in KKR-CPA}

The uni-axial MAE
is obtained as the derivative of the free energy
following Eq.~(\ref{eq::torque_as_the_derivative}).
Inserting Eq.~(\ref{eq::free_energy})
for the free energy,
we get the following~\cite{julie_2006}.
\begin{equation}
T^{(\hat{\mathbf{n}})}_{\theta}=-\frac{\partial}
{\partial\theta}
\sum_{i}
\int P_{i}^{(\hat{\mathbf n})}(\hat{\mathbf e}_{i})
\left<
\Omega^{(\hat{\mathbf n})}
\right>_{\hat{\mathbf e}_{i}}
d\,
\hat{\mathbf e}_{i}
\label{Eq::MAE}
\end{equation}
This is our working formula
to produce the results for
the finite-temperature MAE of YCo$_5$ which we now go on to describe.

\subsection{
Specifics with the case of YCo$_5$}
\label{method::details}

The past decade has seen the successful application of DLM theory to
L1$_{0}$-FePt alloy~\cite{julie_2004, julie_2006}
which is a uniaxial ferromagnet and
the magnetism is carried largely by $3d$ and $5d$ electrons.
Since YCo$_5$ involves mostly $3d$-electrons for its magnetism,
success of DLM description
on the same level as was achieved for
L1$_{0}$-FePt is expected provided that the assumption of localized
moment on the magnetic atoms works well.
Some care must be taken in the application
of DLM for Co-based magnets
since the trend
among Fe, Co, and Ni~\cite{hubbard_1981} shows that
the assumption of localized moment works well for Fe,
faces challenge for Ni, and Co sits somewhere in-between.
Stability of local moments on the Co sublattices in the present case
is discussed
in Sec.~\ref{sec::local_moment_robustness}. Indeed we will
see below some inherent underestimate near the Curie point
in our calculations
which might indicate some fragile nature of the local moments in YCo$_5$.
This problem can be cured by extending the calculation to
that based on the non-local CPA~\cite{PRB_nl_CPA,julie_2014}. For now
we are mostly concerned with the lowest temperature properties and
the middle-temperature range of $T\stackrel{<}{\sim} 600~\mbox{[K]}$
which is well below
the Curie temperature at $T_{\rm Curie}=920~\mbox{[K]}$~\cite{neutron_1980}.
So we proceed with the single-site theory within the scope of the present
project.

\subsubsection{Crystal structure}

\begin{figure}
\begin{center}
\begin{tabular}{l}
(a)\\
\scalebox{0.7}{\includegraphics{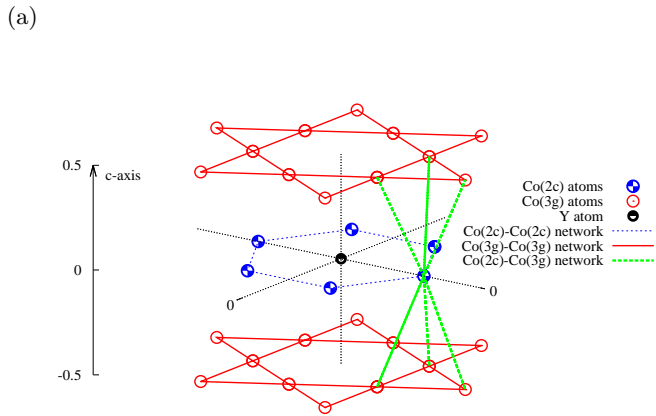}}\\
(b)\\
\scalebox{0.7}{\includegraphics{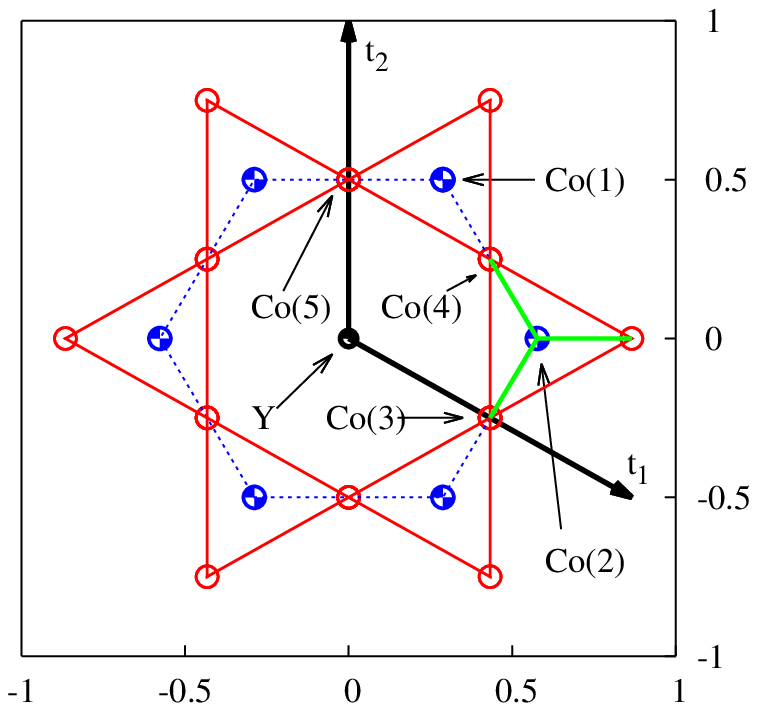}}\\
\end{tabular}
\end{center}
\caption{\label{crystal_structure_cacu5} (Color online) Crystal structure
of YCo$_5$: (a) bird's eye view
and (b) top view seen along the $c$-axis.
Nearest-neighbor pairs
in the Co(2c)-Co(3g) network are only partially drawn for the illustration.
Labels for atoms in (b) follow Table~\ref{atomic_positions}. 
}
\end{figure}
We take the experimental crystal structure of YCo$_5$
with the space group No. 191
$a=4.948$~[\AA] and $c=3.975$~[\AA]~\cite{neutron_1980}. The unit cell
consists of one formula unit: Y, Co(2c), and Co(3g) sublattices.
The atomic
configuration in the unit cell
is shown in Fig.~\ref{crystal_structure_cacu5}~(a)
as a bird's eye view. The top view along the $c$-axis
is shown in Fig.~\ref{crystal_structure_cacu5}~(b)
to illustrate the relative positions of
the Co(2c)-atoms and the Co(3g)-atoms.
Within the layer along the $ab$-plane, it is seen
that nearest-neighbor Co(3g)-atoms form a kagom\'{e} lattice.
Between those kagom\'{e}
layers, Co(2c)-atoms sit right on the center of Co(3g)-triangles
to form a hexagonal lattice on the same $ab$-plane layer as Y-atoms.
The ratio $c/a\simeq 0.80$ means that
the tetrahedron formed by the Co(3g)-triangle
and the Co(2c)-atom is close to the regular one.
So the Co(2c)-Co(3g) atomic distance is
almost equal to Co(3g)-Co(3g) distance which is $0.5a$. On the other
hand, Co(2c)-Co(2c) distance which is $a/\sqrt{3}\simeq 0.58a$ spans
the longest distance among the nearest-neighbor Co atom pairs.
We set the position of each atom and the direction of the translation vectors
as shown in Table~\ref{atomic_positions} in our calculation.
The labeling
scheme for the atoms and the directions
is illustrated in Fig.~\ref{crystal_structure_cacu5}~(b).
\begin{table}
\begin{tabular}{ccc}\hline
sublattice & atom & atomic coordinates \\ \hline
\multicolumn{2}{c}{Y} & $(0,0,0)$ \\ \hline
Co(2c) & Co(1) & $(1/(2\sqrt{3}),1/2,0)$  \\ \hline
& Co(2) & $(1/\sqrt{3},0,0)$ \\ \hline
Co(3g) & Co(3) & $(\sqrt{3}/4,-1/4,1/2)$ \\ \hline
& Co(4) & $(\sqrt{3}/4,1/4,1/2)$ \\ \hline
& Co(5) & $(0,1/2,1/2)$ \\ \hline
\end{tabular}
\begin{tabular}{cc}\hline
\multicolumn{2}{c}{primitive translation vectors}  \\ \hline
${\mathbf t}_{1}$ & $(\sqrt{3}/2,-1/2,0)$ \\ \hline
${\mathbf t}_{2}$ & $(0,1,0)$ \\ \hline
${\mathbf t}_{3}$ &  $(0,0,1)$ \\ \hline
\end{tabular}
\caption{\label{atomic_positions} Set-up of
the unit cell of YCo$_5$ in the present calculations.
}
\end{table}

\subsubsection{Robustness of local moments in YCo$_5$}
\label{sec::local_moment_robustness}

\begin{figure}
\begin{center}
\scalebox{0.8}{\includegraphics{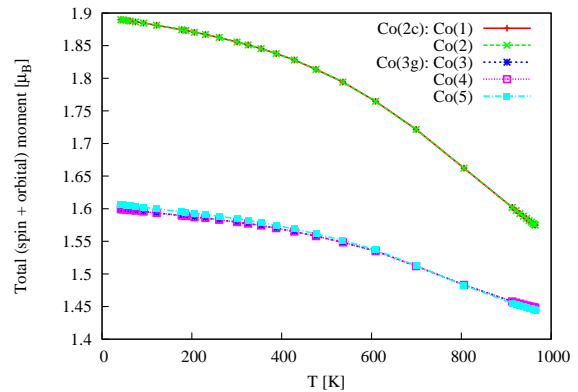}
}
\end{center}
\caption{\label{lsm_total} (Color online) Calculated temperature dependence of
the total (including both spin and orbital) local moment for each magnetic sublattice in YCo$_5$.}
\end{figure}
We inspect the validity of DLM approach at finite temperatures
in the present calculation. Calculated temperature dependence of each
of local moments in YCo$_5$ are shown in Fig.~\ref{lsm_total}.
The DLM picture of magnetism in metals
at finite temperature is based
on the assumption that there are
some aspects of the interacting electrons
of a material that vary slowly
in comparison with faster degrees of freedom.
These are captured in terms of the orientational unit vectors,
$\{\hat{\mathbf e}\}$,
the transverse part of the magnetic fluctuations.
The electronic grand
potential $\Omega^{(\hat{\mathbf n})}(\{\hat{\mathbf e}\})$ is
in principle available from spin-resolved DFT (SDFT)
generalised
for the non-collinear magnetic profiles
which are labelled by $\{\hat{\mathbf e}\}$~\cite{gyorffy_1985}.
Within the tenets of SDFT the charge and magnetisation densities
depend on these configurations,
i.e. $\rho({\bf r},\{\hat{\mathbf e}\})$, $\vec{M}({\bf r},\{\hat{\mathbf e}\})$.
In the space around an atom at a site $i$ the magnetisation
is constrained to follow
the orientation $\hat{\mathbf e}_i$
there but its magnitude $\mu({\bf r})$ can
depend on the surrounding orientational environment, 
$\mu({\bf r})= \mu({\bf r},\{\hat{\mathbf e}\})$.
In many of the materials where the DLM theory
has been successfully applied,
the magnitudes are found to be rather
insensitive to the environment and
can be modelled accurately as rigid local moments.
In other materials, e.g. Ni,~\cite{Staunton+Gyorffy1992},
the magnetism is driven by the coupling of the magnitudes,
(longitudinal magnetic fluctuations) with these transverse modes.
When there is short-range aligning of the orientations
or long range magnetic order, a finite magnetisation magnitude
can establish; whereas in environments where the orientations
differ significantly over short distances, the magnitudes shrink.

Figure~\ref{lsm_total}
indicates the extent of this effect in YCo$_5$.
We have started with a SDFT calculation
of the charge and magnetisation density
$\rho({\bf r})$ and $\vec{M}({\bf r})$
for the $T=0$K ferromagnetic state,
$\{\hat{\mathbf e}= \hat{\mathbf n}\}$ and
find the magnetisation magnitude
per site is 1.9 and 1.6 $\mu_B$
for the Co(2c) and Co(3g) sites respectively.
When we use the scf potentials
that these produce in a frozen potential approximation
in our DLM theory for increasing temperatures
when the moment orientations become disordered,
the average magnetisation magnitude per site output
from this calculation diminishes.
For temperatures up to $\sim 600$~[K],
this decrease is slight,
thereafter more significant as seen in the figures.
Up to $600~\mbox{[K]}$
therefore the DLM theory as applied is adequate
and we can examine the effects of transverse
fluctuations on the MAE that it describes.
For higher temperatures however,
the effects of
short-ranged correlations
among the $\hat{\mathbf e}$ orientation
(via the use of the non-local CPA
for example~\cite{PRB_nl_CPA,julie_2014}) and
longitudinal fluctuations~\cite{longitudinal}
need to be addressed for a more complete picture.
Comprehensive analyses on bcc-Fe, where the robustness
of the local moments are established,
can be found in a recent work~\cite{1403.2904}.

Further details about the calculations are described
in Appendix~\ref{sec::calc_details}.

\section{Results}
\label{pure}

\begin{figure}
\begin{center}
\scalebox{0.8}{\includegraphics{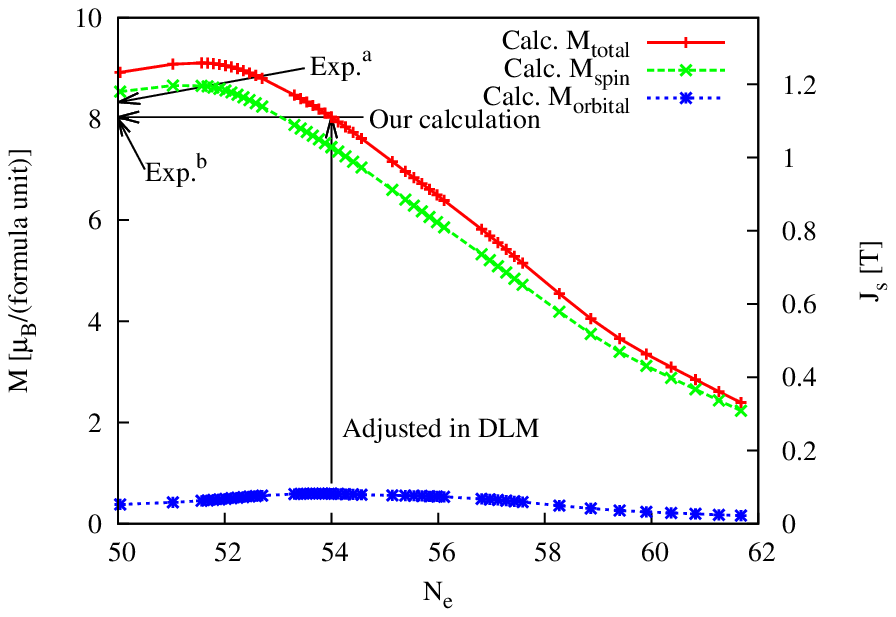}}
\end{center}
$^a$ $M=8.33$~[$\mu_{\rm B}$] at liquid-helium temperature from Ref.~\onlinecite{givord_1981}.\\
$^b$ $M=7.99$~[$\mu_{\rm B}$] at room temperature
from Ref.~\onlinecite{neutron_1980}.
\caption{\label{T_0_pure_mag}
(Color online) Calculated
magnetization of YCo$_5$ plotted as a function
of the number of electrons
near the ground state at $m=0.967$.
Our calculated result
8.03~[$\mu_{\rm B}$] at around $T=100$~[K]
falls onto the experimental numbers
within 2 digits. For the closeness of such data
to the ground state, see Fig.~\ref{Mag_pure}.}
\end{figure}

We start with the calculated MAE near the ground state
to demonstrate its sensitivity to the filling of the electrons
in Fig.~\ref{T_0_pure}. Corresponding data for the magnetization
is shown in Fig.~\ref{T_0_pure_mag}. We
see that even the sign of the MAE sensitively changes
depending on a fraction in the filling of electrons,
which is
reasonable
in the physics of magnetic anisotropy
where adding/removing one electron
in an electronic cloud that points to a particular real-space direction
in a given crystal-field environment
indeed affect which direction the magnetic moment should prefer
under the presence of SOI.
So the challenge is how precisely
we can pin-point the electron number right onto the realistic one,
on which details are given in Sec.~\ref{sec::fixing_the_filling}.
Fixing the electron number
by a fine-tuning of the chemical potential in the DLM runs,
we obtain the temperature dependence of MAE and magnetization.
The temperature dependence of the anisotropy field
is deduced from them. These data
for the bulk are shown in Sec.~\ref{sec::bulk}. 
Then we inspect how our calculated MAE scales with respect to the calculated
magnetization, resolving into sublattices and temperature ranges
in Sec.~\ref{results::scaling_mae_vs_mag}.
Comparison to the
single ion anisotropy model analyses by Callen and Callen~\cite{callencallen}
and more recent ones~\cite{julie_2004,oleg_epl} is carried out to help uncover 
the key mechanisms.

\subsection{Bulk observables}
\label{sec::bulk}

\begin{figure}
\begin{center}
\scalebox{0.8}{\includegraphics{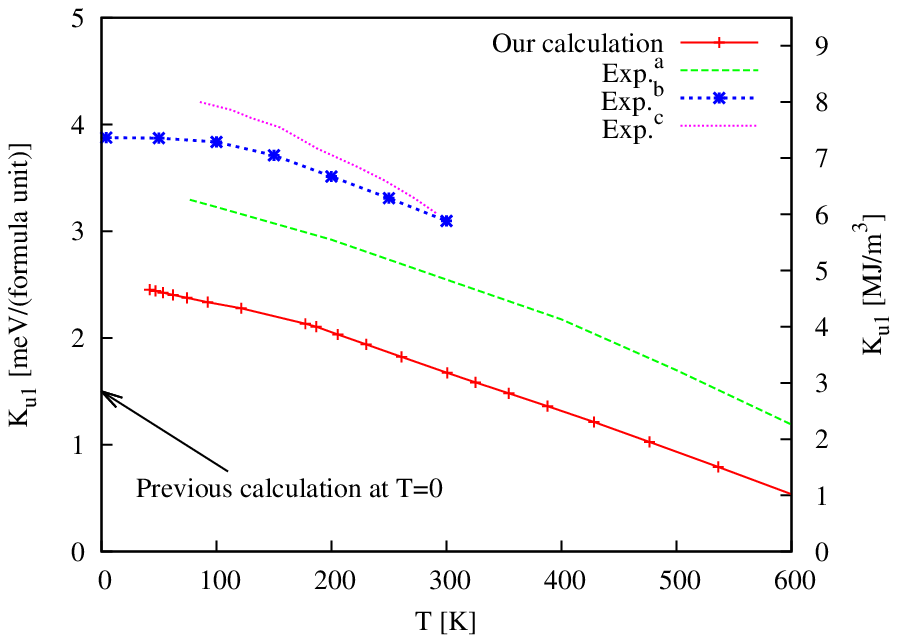}
}
\end{center}
$^a$ Ref.~\onlinecite{buschow_1974} for $K_{\rm u1}$.
$^b$ Ref.~\onlinecite{givord_1981} for $K_{\rm u1}$.\\
$^c$ Ref.~\onlinecite{Pr_doping_1988} for $K$, with $K_{\rm u1}$ dominating.
\caption{\label{MAE_pure} (Color online)
Calculated temperature dependence of
magnetic anisotropy energy for YCo$_5$.
Previous calculation result at $T=0$ is taken from Ref.~\onlinecite{igor_2003}.
}
\end{figure}
Calculated temperature dependence of MAE
for YCo$_5$ by DLM
at the fixed
correct valence-electron number
is shown
in Fig.~\ref{MAE_pure}.
We have a systematic underestimate of MAE
for the overall temperature range
as compared to
the experimental results found in the literature.
However the qualitative temperature dependence
is
well reproduced.
\begin{figure}
\begin{center}
\scalebox{0.8}{\includegraphics{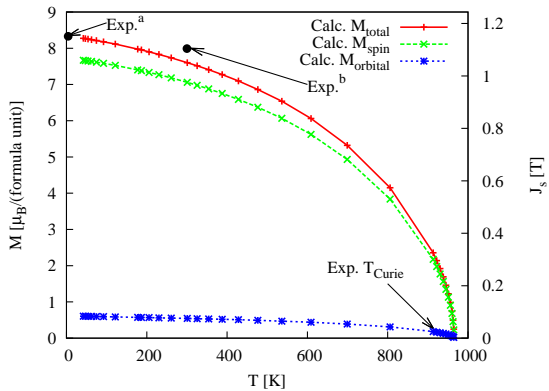}
}
\end{center}
$^a$ Ref.~\onlinecite{givord_1981} $^b$ Ref.~\onlinecite{neutron_1980}
\caption{\label{Mag_pure} (Color online) Calculated temperature dependence of
magnetization
for YCo$_5$.
Experimental Curie temperature, 920~[K],
is taken from Ref.~\onlinecite{neutron_1980}.
}
\end{figure}
In the similar manner to MAE,
calculated temperature dependence of magnetization at
the fixed valence-band filling
is shown in
Fig.~\ref{Mag_pure}. Fitting the calculated
temperature dependence of magnetization 
as it decreases to zero
in the temperature range
$T>900$~[K]
to the following
relation
\[
M(T)= A_{\rm M}\left|T-T_{\rm Curie}\right|^{\beta},
\]
we get $T_{\rm Curie}=965$~[K], $\beta=0.50$,
and $A_{\rm M}=0.33$.
The critical exponent
falls onto
the mean-field value $\beta=1/2$ within the numerical accuracy
as it should, which gives a check of the data.
Our {\it ab initio} result for the Curie temperature
compares with the experimental Curie temperature $T_{\rm Curie,exp}=920$~[K]~\cite{neutron_1980}
within the deviation of 5\%.
A mean-field result on the transition temperature
typically gives an overestimate by $O(10)\%$ as was observed for
L1$_{0}$-FePt alloy calculated by exactly the same method~\cite{julie_2004} as
the one we utilize here; the apparent
absence of such an
overestimate might rather
indicate the
fragility of the local moments close to $T_{\rm Curie}$.

\begin{figure}
\begin{center}
\scalebox{0.8}{\includegraphics{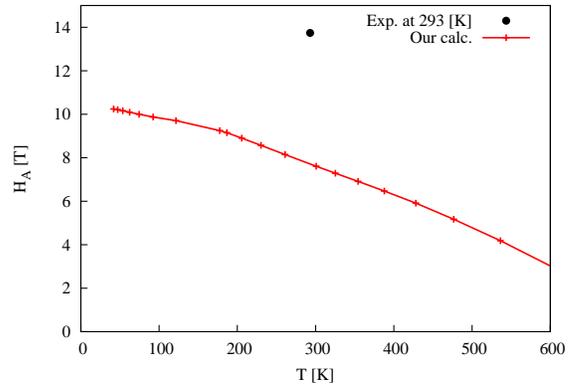}}
\end{center}
\caption{\label{anisotropy_field_pure} (Color online)
Temperature dependence of
the anisotropy field
for YCo$_5$ as deduced from the numbers presented
in Figs.~\ref{MAE_pure}~and~\ref{Mag_pure},
following Eq.~(\ref{anisotropyfield}).
The experimental data at 293 [K]
is taken from Ref.~\onlinecite{Pr_doping_1988}.
}
\end{figure}

We obtain {\it the}
{\it ab initio} result
for the temperature dependence of the anisotropy field $H_{\rm A}$
as shown in Fig.~\ref{anisotropy_field_pure}~(a) following the simple
coherent magnetization rotation
picture,
\begin{equation}
K_{\rm u1}=\frac{1}{2}M H_{\rm A},
\label{anisotropyfield}
\end{equation}
with $K_{\rm u1}$ and $M$
being the
calculated
uniaxial MAE in Fig.~\ref{MAE_pure} and magnetization in Fig.~\ref{Mag_pure},
respectively.
The underestimate by 30\%
of $K_{\rm u1}$
on the left-hand side of Eq.~(\ref{anisotropyfield})
leads to a similar underestimate
of $H_{\rm A}$.

\subsection{Scaling MAE and Magnetization}
\label{results::scaling_mae_vs_mag}

Scaling relation concerning
MAE with respect to magnetization
\begin{equation}
K_{\rm u1}(T)=A \left[M(T)/M(0)\right]^{\alpha}
\label{callen}
\end{equation}
has been discussed for magnetic materials
of which the representative one was established by Callen and Callen~\cite{callencallen}
on the basis of single-ion theory.
Their exponent $\alpha=l(l+1)/2$ at low temperatures,
where $l$ is the order of the expansion in terms of the spherical harmonics,
has been challenged in some materials~\cite{julie_2004,oleg_epl}.
Callen-Callen argument for a ferromagnet
with the uni-axial anisotropy predicts an exponent $\alpha=3$ with $l=2$.
Here we inspect what sort of scaling relation holds for our target magnet YCo$_5$.
YCo$_5$ is an intermetallic compound
and the single-ion theory is not directly applicable to it. On top of that,
the particular crystal structure of YCo$_5$
comes with multiple magnetic sublattices in the unit cell.
The bulk scaling between
MAE and magnetization
might not be exactly the same as sublattice-resolved scaling
concerning the fact that each sublattice is exposed to its own
crystal-field environment, depending on
its own characteristic temperature dependence.

\begin{figure}
\begin{center}
\scalebox{0.8}{\includegraphics{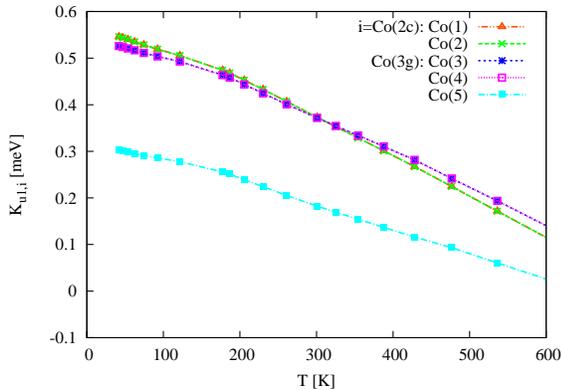}
}
\end{center}
\caption{\label{local_moment_resolved_mag_and_mae} (Color online)
Calculated 
temperature dependence of local-moment-resolved MAE.}
\end{figure}

\begin{figure}
\begin{center}
\scalebox{0.8}{\includegraphics{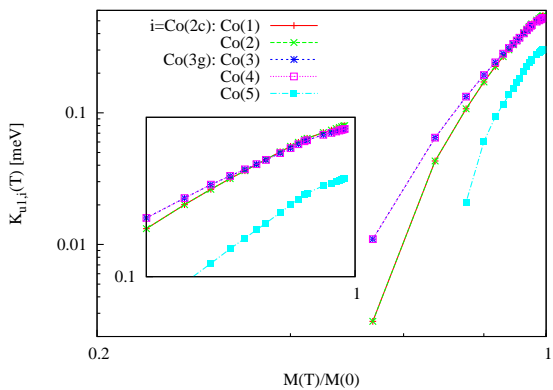}
}
\end{center}
\caption{\label{local_moment_resolved_mag_vs_mae} (Color online)
Calculated MAE
plotted as a function of the order parameter, $m\equiv M(T)/M(0)$, for each local moment $i$.
The inset is a zoomed-in plot in region $M(T)/M(0)\stackrel{<}{\sim} 1$.}
\end{figure}

\begin{figure}
\begin{center}
\scalebox{0.8}{\includegraphics{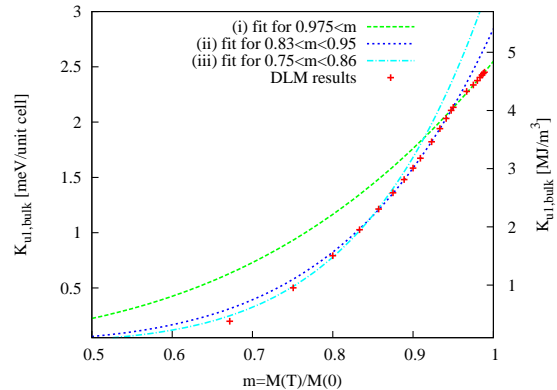}
}
\end{center}
\caption{\label{scaling_mae_vs_mag} 
(Color online)
Scaling of the bulk MAE, $K_{\rm u1,bulk}(T)$,
with respect to the order parameter $m=M(T)/M(0)$. The scaling exponent $\alpha$
in $K_{\rm u1,bulk}(T)=A m^{\alpha}$ depends on the temperature-dependent
scaling region: i) $\alpha=3.5$ for
$m>0.975$ which corresponds to $41~\mbox{[K]}<T<75~\mbox{[K]}$,
ii) $\alpha=5.6$ for
$0.83<m<0.95$ which corresponds to $186~\mbox{[K]}<T<477~\mbox{[K]}$,
and iii) $\alpha=6.5$ for
$0.75<m<0.86$ which corresponds to $428~\mbox{[K]}<T<610~\mbox{[K]}$.
}
\end{figure}

We start with the temperature dependence of
local-moment resolved
MAE as shown in Fig.~\ref{local_moment_resolved_mag_and_mae}.
Exceptional behavior seen for Co(5) in the Co(3g)-sublattice
originates
in the effect of SOI with the magnetization
direction ${\mathbf n}=(1,0,1)$.
Out of the data in Fig.~\ref{local_moment_resolved_mag_and_mae},
the sublattice-resolved MAE as a function of the order parameter
$m=M(T)/M(0)$
is plotted in Fig.~\ref{local_moment_resolved_mag_vs_mae} to which
we apply the scaling relation Eq.~(\ref{callen}) for each of the
contributions from the magnetic sublattices as well as for
the bulk data.
The bulk scaling analysis is shown
representatively
in Fig.~\ref{scaling_mae_vs_mag}.
Experimental bulk data is taken from the past literature~\cite{givord_1981}.
For the temperature range where the reliable experimental data set
seems to be available,
we tabulate the extracted
scaling exponents in Table~\ref{microscopic_scaling_results}.
\begin{table}
\begin{center}
\begin{tabular}{c||ccc} \hline
& $T=177$~[K]& 186~[K]& 300~[K]  \\ \hline
 Co(1)/Co(2) & \--- &
\multicolumn{2}{|c}{$\alpha=5.6$}  \\ \hline
 Co(3)/Co(4) & \--- &
\multicolumn{2}{|c}{$\alpha=4.9$}   \\ \hline
 Co(5) &  \multicolumn{3}{c}{$\alpha=7.7$} \\ \hline\hline
Bulk (Our calc.) & \--- &
\multicolumn{2}{|c}{$\alpha=5.6$}  \\ \hline
Bulk (Exp.$^a$) & 
\multicolumn{3}{c}{$\alpha=6.1$ for $150~\mbox{[K]}\le T\le 300~\mbox{[K]}$}
 \\ \hline
\end{tabular}\\
$^a$ Ref.~\onlinecite{givord_1981}
\end{center}
\caption{\label{microscopic_scaling_results}
Focus temperature range and the extracted scaling exponents
$\alpha$
in $K_{\rm u1}\propto m^{\alpha}$ for each local moment
on the magnetic sublattice in the unit cell of YCo$_5$ and for the bulk.
}
\end{table}
The scaling
exponents do depend on the magnetic sublattices.
We find that the bulk scaling exponent is observed only in a phenomenological way
after summing up the contribution of each magnetic atoms in the unit cell.
The overall calculated
trend from the high-temperature side with larger $\alpha$ to
the low-temperature side with smaller $\alpha$ is shared by all of the sublattice
and the bulk.
The bulk scaling
exponent around the room temperature is calculated to be $\alpha\sim 5.6$
which reasonably compares with the experimental exponent $\alpha\sim 6.1$
that was extracted in the temperature region including the room temperature.

We note that our calculated data 
in Fig.~\ref{scaling_mae_vs_mag}
shifts from experimental data
on the $m$ axis when compared at the same temperatures.
Our calculated data for $m>0.96$ are for below the liquid-nitrogen
temperatures while for the experimental data~\cite{givord_1981} the room-temperature
data is already in the parameter range $m>0.96$.
That
deviation could be related to the inherent underestimate
in our calculations.

\section{Discussions}
\label{sec::coercivity}

We discuss the strategy
to improve the persistence of coercivity against temperature decay.
It has been known that
the coercivity is approximately proportional to
the anisotropy field $H_{\rm A}$~\cite{hirosawa_1986_jmmm,kronmuller_1988}.
A significant part of the reduction factor $H_{\rm c}/H_{\rm A}$
originates
from
the microstructure
which is beyond the scope of
the bulk electronic structure
considerations.
Now our interest lies in how to improve $H_{\rm A}(T)$
or equivalently
$K_{\rm u1}(T)$ for high $T$.
We propose a scheme
to exploit
the peak position of MAE
in Fig.~\ref{T_0_pure}
and how
the electrons are thermally excited
from the ground state
following the Fermi-Dirac distribution.
The idea is to let the thermally excited electrons
populate the targeted electronic state which lies
below
the peak
in Fig.~\ref{T_0_pure} to enhance MCA.
For that, we artificially
lower
the chemical potential
to make some place for thermally excited electrons to populate
at high temperatures.
Then we sacrifice the 
optimal values of the MCA at
low temperatures but
we can extract
the 
optimal
anisotropy field over the operating
temperature ranges.

\begin{figure}
\scalebox{0.8}{\includegraphics{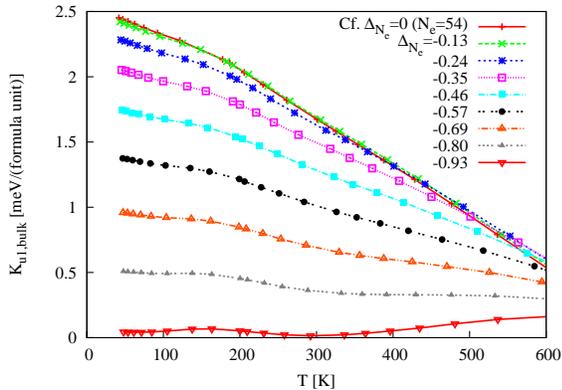}
}
\caption{\label{finite_T_tuning_K_and_M_resolved} (Color online)
Calculated
temperature-dependence of
MAE
for the
hole-doped YCo$_5$.}
\end{figure}

The results for the calculated temperature dependence
of the
MAE
in the hole-doped YCo$_5$'s
is shown in Fig.~\ref{finite_T_tuning_K_and_M_resolved}.
Hole
doping is implemented by artificially lowering the Fermi level
in DLM runs for YCo$_5$.
Measuring the position of the lowered Fermi level,
$E_{\rm F,doped}$,
from the undoped one, $E_{\rm F,undoped}$,
in units of Kelvins
helps us
to figure out around which temperature range
the thermally
excited
electrons would reach the peak position in Fig.~\ref{T_0_pure}.
We found that
for the reduction range $150\mbox{~[K]}\stackrel{<}{\sim}
\Delta_{E_{\rm F}}\stackrel{<}{\sim} 300\mbox{~[K]}$,
with $\Delta_{E_{\rm F}}\equiv
(E_{\rm F,doped} - E_{\rm F,undoped})$,
some enhancement of the MAE
around $T\sim 500\mbox{ [K]}$ is numerically
observed.
The data with $\Delta_{E_{\rm F}}=314$~[K]
and $471$~[K] were extracted with the corresponding reduction in the electron number being
$\Delta N_e=-0.24$ and $-0.35$, respectively, as the main message
of this paper in Fig.~\ref{main_message},
where MAE is expressed
in terms of the anisotropy field following Eq.~(\ref{anisotropyfield}).

Remarkably, non-monotonic temperature dependence of $H_{\rm A}$ is found
for the computational ``sample'' in the low-coercivity region
with the electron number reduced by around 0.9 in
Fig.~\ref{finite_T_tuning_K_and_M_resolved}.
This non-monotonic temperature temperature dependence of MAE
is reminiscent of what has been known
experimentally for
Y$_{2}$Fe$_{14}$B since 1986,
where the temperature-enhancement in MAE
in the order of 0.2-0.3 [MJ/m$^3$] is observed around $T\sim 300$~[K]~\cite{hirosawa_1986}.
The common factor among Y$_{2}$Fe$_{14}$B and YCo$_5$ is
that $3d$-electron contribution in the MCA of rare-earth magnets is extracted out.
The non-monotonicity in our finite-temperature data for
the chemical-potential controlled YCo$_5$
comes from the following two factors:
a)
thermal population of the particular electronic states
that corresponds to the peak of MAE in Fig.~\ref{T_0_pure},
and b) various contributions at finite temperatures
from multiple sublattices in the unit cell.
We speculate that
analogous physics
may be at work in Y$_{2}$Fe$_{14}$B
which
would require
more extensive
computations than those for the present project.

We note that the Curie temperature shifts
as the chemical potential is reduced.
The temperature dependence
of magnetization shifts rather
monotonically with respect to the artificially
manipulated chemical potential while
$K_{\rm u1}(T)$
behaves non-monotonically, 
and
the high-temperature enhancement seen
in Figs.~\ref{main_message}~and~\ref{finite_T_tuning_K_and_M_resolved} is
to be considered as a finite-temperature
reflection of the non-monotonic ground-state
behavior in Fig.~\ref{T_0_pure} rather than a reflection
of the plain shift of the Curie temperature.

\section{Conclusions}
\label{conc}

Temperature dependence of MAE and magnetization for YCo$_5$ has been
calculated from first principles within the assumption of local moments
to give the
qualitative agreement with the experimental data. 
We have proposed a way to improve the high-temperature coercivity
by hole doping as was demonstrated by calculations
with the filling controlled by the chemical potential manipulation.
That can be implemented as applying the gate voltage for the case
of a thin film sample, or
alloying
Co with some other elements
to slightly reduce the 
valence
electron number in the bulk.
Since Co is an expensive element,
replacing it with
some cheaper ingredients with the improved coercivity in the practical operation
temperature range is quite welcome.
Thus we have proposed a way to improve the practical utility of YCo$_5$
from
the perspective of the electronic structure
and statistical physics.

\begin{acknowledgments}
This work is supported by
the Elements Strategy Initiative Center for Magnetic Materials (ESICMM)
under the outsourcing project of the Ministry of Education,
Culture, Sports, Science and Technology (MEXT), Japan.
Support is acknowledged from the EPSRC~(UK)~grant~EP/J006750/1~(J.B.S.~and
~R.B.).
Overall guidance given by T.~Miyake~and~H.~Akai
is gratefully acknowledged.
MM thanks ESICMM people and Y.~Yamaji
for helpful discussions,
L.~Petit and D.~Paudyal for fruitful
interactions during KKR workshop in Warwick in July 2013,
O.~N.~Mryasov for discussions
during his visit to ESICMM, NIMS in July-August 2013,
and
K.~Hono for
the overall push to the present project.
Numerical computations were performed on ``Minerva''
at University of Warwick and SGI Altix at National Institute for Materials Science.
\end{acknowledgments}

\appendix

\section{Details of the calculations for YCo$_5$}
\label{sec::calc_details}

\subsection{Adjustment of the valence electron number}
\label{sec::adjusting_the_filling_by_2_steps}

The overall calculation consists of two steps:
\begin{enumerate}
\item The first stage is to generate the grand potentials
for each of the assumed local moment configurations
by a scalar-relativistic KKR
calculation.
We do a spin-polarized 
calculation for YCo$_5$
as parametrized by Vosko {\it et al.}~\cite{vosko_1980}
to reach the ferromagnetic ground state
and yield the potentials for each local moment
on the magnetic sublattices.
\item Then the second stage
consists of solving the fully relativistic equation of state under the potentials
generated by the electronic structure calculation in step 1
and the direction of magnetization is fixed to be ${\mathbf n}=(1,0,1)$
with the motivation to address MAE using the torque-based formula in Eq.~(\ref{Eq::MAE}).
Solutions are obtained
so that Eqs.~(\ref{Weiss_field_expressed_as_})
and~(\ref{PDF}) are solved self-consistently
to yield
the result on MAE, magnetization, and the temperature.
\end{enumerate}
In these KKR-based calculations
we set the cutoff order in the expansion of spherical harmonics
$l_{\rm max}=3$ all through the above two steps.

In step 1,
we follow the
atomic-sphere approximation (ASA) considering the deformed shape
of the unit cell to avoid the complication caused by overlapping muffin-tin spheres
or too large a interstitial region. We have confirmed that this ASA construction is adequate
from full potential (FP)-linear muffin tin orbital (LMTO) calculations where
we find the ASA and FP-LMTO densities of states for the $T=0$K cases to compare well.

\begin{figure}
\begin{center}
\scalebox{0.8}{\includegraphics{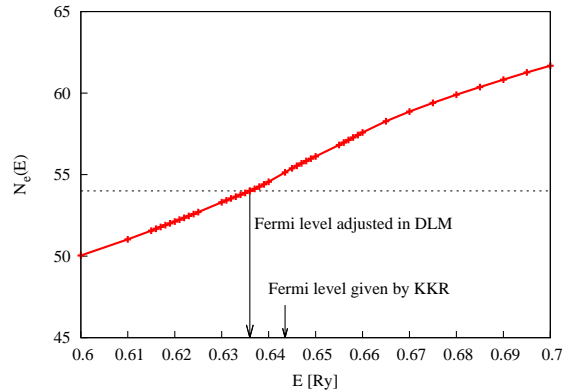}}\\
\end{center}
\caption{\label{T_0_TOTDOS_pure} (Color online)
Calculated Fermi-level dependence of
number of electrons
for pure YCo$_5$
near the ground state at $m=0.967$ [$m=L(\beta h)$ with $\beta h=30$
as in Eq.~(\ref{Langevin})].
Equation of state is solved
with the artificially given Fermi level
to adjust the valence electron number exactly to the number
as set up by the calculation.
}
\end{figure}

Having difference in the treatment
of the SOI between 
steps 1 and 2,
adjustment of the Fermi level needs to be done to exactly fix the number of electrons,
which is demonstrated in Fig.~\ref{T_0_TOTDOS_pure}.
The adjustment
is done in the order
of $O(0.01)~\mbox{[Ry]}\sim O(0.001)~\mbox{[eV]}$. We describe
the practical procedure for this in Sec.~\ref{sec::fixing_the_filling}.

\subsection{Achieving the self-consistency in DLM with multiple magnetic sublattices in the unit cell}
\label{ultimate_scf}

Further iterations within DLM is taken
for YCo$_5$ to achieve the self-consistency of the calculation
to incorporate the effects
of multiple mean-fields that depend on the magnetic sublattice
within the unit cell. The target material YCo$_5$
has at least two different magnetic sublattices,
Co(2c) and Co(3g), and the magnetic moments on them can be different from each other
as was found in the neutron scattering experiments~\cite{neutron_1980}.
The Weiss fields are obtained
from Eqs.~(\ref{Weiss_field_expressed_as_})~and~(\ref{PDF})
solved self-consistently for the sites
on the sub-lattice where there are local moments residing.
We proceed as follows.
\begin{enumerate}
\item We start with a run
with $\beta h$ uniformly set over all of the local moments in the unit cell
so that $m_i = L(\beta h)$.
We refer to this run the zero-th round of the correction process.

\item With the output of the zero-th round,
the input $\beta h$'s are re-set by using the output Weiss fields produced by the calculation from Eq~(\ref{Weiss_field_expressed_as_}).
\begin{equation}
\beta=(\beta h)_{i}/h_{i}
\end{equation}
for all $i$'s. We note that
the left-hand side, inverse temperature,
is sublattice-independent. This imposes the self-consistency
condition: we fix the input $\beta h$ for Co(2c)-sublattice and adjust
$\beta h$ for other local moments that sit on the other sublattices
according to the following relation:
\begin{equation}
\frac{(\beta h)_{\rm Co(3g)}}{h_{\rm Co(3g)}}
\equiv
\frac{(\beta h)_{\rm Co(2c)}}{h_{\rm Co(2c)}}
\label{EoS_self_consistency}
\end{equation}
Using the output of the zero-th round for $h$'s,
the input numbers for
$(\beta h)_{\rm Co(3g)}=h_{\rm Co(3g)}\times
(\beta h)_{\rm Co(2c)}/h_{\rm Co(2c)}
$ is used for the first round.
\item The process is iterated, making the 2nd, 3rd,\ldots rounds in the correction process,
until the input $S^{1}$ and the output $S^{1}$'s are identical within the numerical accuracy.

\end{enumerate}

We have observed that for $(\beta h)_{\rm Co(2c)}\ge 10$
the convergence within 4 digits
for $h$'s are quickly
reached typically after 3 iterations while for $(\beta h)_{\rm Co(2c)}\stackrel{>}{\sim} 0.1$
the convergence takes around 10 iterations. The iteration can run into a limit cycle
of a few periods beyond the 5th digit of $\beta h$.
At such stage of the iteration,
we regard that the convergence was reached
within the numerical precision of 4 digits for $S^1$
which we believe is sufficient.

\subsection{
Determining and varying the chemical potential in the DLM}
\label{sec::fixing_the_filling}

In addressing magnetic hardness from first principles,
magnetic anisotropy energy (MAE) needs to be calculated
as precisely as possible.
The behavior of the calculated
MAE as a function of the valence electron number
for YCo$_{5}$
is shown in Fig.~\ref{T_0_pure}. This data is obtained
in a series of DLM runs by
artificially modulating
the chemical potential near the ground state.
Here the data near the ground state means
the order parameter is fixed to be $m=0.967=L(\beta h)$ with $\beta h=30$.
MAE has the strong dependence on the electron filling
and only after fixing the electron number
by a fine-tuning of the chemical potential in the DLM run,
we can discuss the temperature dependence of MAE and magnetization
from first principles: here we fix the valence electron number down to 4 digits
by manual tuning the chemical potential. The fine-tuning follows
the Newton's method in spirit, where we
keep on shifting the chemical potential by $\Delta\mu$
with
\[
\Delta \mu=(N_{e,{\rm desired}}-N_e)/n(\mu)
\]
where
$N_e$ is the electron number given
as $N^{(\hat{\mathbf{n}})}(\mu)$
and $n(\mu)$
is the density of states
at $\mu$,
until the desired electron number $N_{\rm desired}=54$ is numerically
reached
by
iterating
the calculations
with the shifted chemical potential ($\mu+\Delta\mu$).
At
this determined
chemical potential and the electron number $N_e=54.00$
near the ground state,
we find the low-temperature MAE
to be 2.3~[meV/unit cell] as shown in Fig.~\ref{T_0_pure}.
Using the experimental lattice constants~\cite{neutron_1980},
this is
equivalent to
4.4~[MJ/m$^3$],
which compares with the experimental result
at liquid nitrogen temperature,
6.3~[MJ/m$^3$]~\cite{buschow_1974},
with an underestimate of 30\%.

The above fine-tuning of the chemical potential
also depends
on the temperature. At each data point on the temperature axis,
the valence electron number is fixed to be 54.00 within the numerical accuracy
which practically gives a temperature-dependent chemical potential
from 0.63609~[Ry] at the lowest temperatures
to 0.63357~[Ry] at the calculated highest temperature 965~[K].

\subsection{
Magnetization and local moments}
\label{mag_and_local_moments_T_0}

The total magnetization, and its spin part, and the orbital part,
is calculated to be
8.03~[$\mu_{\rm B}$],
7.44~[$\mu_{\rm B}$],  and 0.59~[$\mu_{\rm B}$], respectively,
as
shown in Fig.~\ref{T_0_pure_mag} for low temperature.
The calculated total magnetization
amounts to $J_{\rm s}=1.11$~[T]
using the experimental lattice constants~\cite{neutron_1980}.
These results agree with
the experimental numbers found in the literature
of
7.99~[$\mu_{\rm B}$]~\cite{neutron_1980} or $J_{\rm s}=1.1$~[T]~\cite{ohashi_2012}
within $0.5\%$.

However we note that
this apparent excellent agreement
is actually
realized by a cancellation of overestimates and underestimates,
which is revealed by
inspecting each local moment's contribution to the total magnetization.
At the fixed
filling,
our results are extracted
as summarized in Table.~\ref{T_0_moments_numbers}.
Our calculated local moments on Co(2c)/Co(3g) sublattice
are 
1.88 and 1.59~[$\mu_{\rm B}$]
which compare with the results from neutron experiments
1.77/1.72~[$\mu_{\rm B}$]~\cite{neutron_1980}
within 6-8\%. Those polarizations are
given by DLM in the fully relativistic calculation near the ground state,
based on the potentials generated by
the scalar-relativistic KKR calculation at $T=0$ which
gave the polarization
as 1.48 and 1.52 for Co(2c) and Co(3g), respectively.
In the present local moment approach for YCo$_5$
we have systematic underestimate
of
the orbital moments compared with the neutron measurements.
However, the resolution of the neutron data into spin and orbital
contribution also
involves some subtlety~\cite{saito}. The accurate {\it ab initio} estimation
of the orbital contribution to the magnetization is left for future projects.
\begin{table}
\begin{tabular}{cccc} \hline
& total & spin & orbital \\ \hline
Co(2c) (our calc.) & 1.88 & 1.81 & 0.0667 \\ \hline
Co(2c) (exp.) & 1.77 & 1.31  & 0.46 \\ \hline\hline
Co(3g) (our calc. for Co(3)/Co(4)) & 1.59  & 1.44 & 0.150 \\ \hline
(our calc. for Co(5))  & 1.60 & 1.44 & 0.164 \\ \hline
Co(3g) (exp.) & 1.72 & 1.44 & 0.28 \\ \hline
\end{tabular}
\caption{\label{T_0_moments_numbers} 
Calculated magnitude of local moments in~[$\mu_{\rm B}$].
Experimental numbers are taken from Ref.~\onlinecite{neutron_1980}.
}
\end{table}

\bibliography{myreferences}

\end{document}